\documentclass[lettersize,journal]{IEEEtran}
\usepackage{amsmath,amsfonts}
\usepackage{algorithmic}
\usepackage{algorithm}
\usepackage{array}
\usepackage[caption=false,font=normalsize,labelfont=sf,textfont=sf]{subfig}
\usepackage{textcomp}
\usepackage{stfloats}
\usepackage{url}
\usepackage{verbatim}
\usepackage{graphicx}
\usepackage{cite}
\usepackage{tikz}
\usepackage{xcolor}

\begin{document}

\title{Computation of Characteristic Mode for Regional Structure of Interconnected Metal Bodies}

\author{{Chenbo Shi, Jin Pan, Xin Gu, Shichen Liang and Le Zuo}
\thanks{Manuscript received Mar. 4, 2024; revised Jul. 15, 2024. This work was supported by the Aeronautical Science Fund under Grant ASFC-20220005080001. (\textit{Corresponding author: Jin Pan.})}
\thanks{Chenbo Shi, Jin Pan, Xin Gu and Shichen Liang are with the School of Electronic Science and Engineering, University of Electronic Science and Technology of China, Chengdu 611731 China  (e-mail: chenbo\_shi@163.com; panjin@uestc.edu.cn; xin\_gu04@163.com; lscstu001@163.com).}
\thanks{Le Zuo is with The 29th Research Institute of China Electronics Technology Group Corporation (e-mail: zorro1204@163.com)}
}

\markboth{}%
{Shell \MakeLowercase{\textit{et al.}}: A Sample Article Using IEEEtran.cls for IEEE Journals}


\maketitle

\begin{abstract}
Existing methods for calculating substructure characteristic modes require treating interconnected metal structures as a single entity to ensure current continuity between different metal bodies. However, when these structures are treated as separate entities, existing methods exhibit inaccuracies, affecting the assessment of structural performance. To address this challenge, we propose an enhanced electromagnetic model that enables accurate characteristic mode analysis for regional structures within interconnected metal bodies. Numerical results validate the accuracy of the proposed method, and an antenna design example demonstrates its practical utility.
\end{abstract}


\section{Introduction}
\IEEEPARstart{C}{haracteristic} mode theory (CMT)  is a vital tool in antenna design, providing engineers with profound insights into the intrinsic electromagnetic (EM) properties of radiating structures \cite{ref_review1,ref_review2}. The original CMT was developed to examine radiation or scattering problems of entire structures within free space \cite{ref_TCM_Origin}. However, for many antenna designs, it is essential to understand the intrinsic EM properties of structures in non-free-space backgrounds, such as near biological tissues or antennas mounted on large platforms. To address these needs, an enhanced variant called substructure CMT (SCMT) exhibits has shown potential \cite{ref_STCM_Concept}. SCMT classifies structures into two categories: the key (controllable) and the background structures. It then formulates an equation to extract the characteristic mode of the key structure (i.e., SCM) by capturing the background structure's effects. The core of this technique involves using the Green's function of a specific background space instead of free space for EM modeling. Recently, researchers have generalized this technique to encompass a wide variety of structure types \cite{ref_STCM_Huang1,ref_STCM_Huang2,ref_Zhao,ref_Shi}.

Despite the success in the theoretical and application aspects of SCMT \cite{ref_STCM_Microstrip,ref_STCM_DRA,ref_Guo_Substructure}, inaccuracies have been identified when determining the EM properties of key metal structures connected to a metallic background. As elaborated further in this work, these inaccuracies stem from an artificial discontinuity in the conductive current at the junction boundary between the key and background structures. This discontinuity restricts the application of SCMT in certain antenna designs. For example, it proves ineffective in extracting the characteristic mode for radiative structures electrically connected to a metallic ground and is also limited in handling antennas that incorporate shorting structures.

To broaden the application of SCMT, this article introduces a transition band that enforces the continuity of conductive currents at the junction boundary between key and background structures. This ensures the accurate extraction of the key regional structure's characteristic modes (referred to as ``RCMs'' for distinction) from interconnected metallic bodies. To validate the significance of this method, we begin by highlighting the discrepancies in the EM properties between full-wave simulation and SCM through an illustrative example where a key PEC structure approaches and eventually touches a PEC sphere. In this example, we directly use the analytical Green's function of a PEC sphere to extract SCMs for rigor. Our RCM technique, however, eliminates these discrepancies and demonstrates high accuracy. We then apply it to a dual-band monopolar patch antenna (MPA) detailed in \cite{ref_MPA}, successfully expanding the operating bands from two to three and enhancing the bandwidth. These examples illustrate the significance of the proposed technique in antenna design.

\section{Characteristic Mode Computation for Regional Structure from Interconnected Metal Bodies}
\label{SecII}

\subsection{Demonstrating SCMT Inaccuracies}
\label{SecIIA}

\begin{figure}[!t]
  \centering
  \subfloat[]{\includegraphics[]{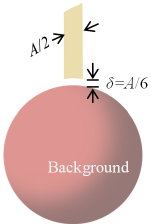}%
  \label{fSheet_sphA}}
  \hfil
  \subfloat[]{\includegraphics[]{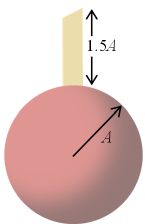}%
  \label{fSheet_sphB}}
  \caption{A PEC sheet near a PEC shpere. (a) case \#1, separated. (b) case \#2, connected.}
  \label{fSheet_sph}
\end{figure}

An illustrative example for assessing the performance of SCMT is presented in Fig. \ref{fSheet_sphA}. In this setup, a PEC sphere of radius $A$ is placed at the origin of the coordinates (denoted as $S_b$), and a PEC sheet (denoted as $S_k$) is suspended above it. The SCM of the PEC sheet is the characteristic mode determined by 
\begin{equation}
  \label{eq1}
  \mathbf{\tilde{Z}J}_{k,n}=\left( 1+j\lambda _n \right) \mathbf{\tilde{R}J}_{k,n}
\end{equation}
Here, $\mathbf{\tilde{Z}}$ is the impedance matrix obtained by discretizing the electric field integral equation on $S_k$ using the Green's function of $S_b$ \cite{ref_Tai_Dyadic} \footnote{The original method employs numerical Green's function \cite{ref_Newman_Overview} to compute SCM,  as described in \cite{ref_STCM_Concept}, where $\tilde{\mathbf{Z}}=\mathbf{Z}_{kk}-\mathbf{Z}_{kb}\mathbf{Z}_{bb}^{-1}\mathbf{Z}_{bk}$. In this expression, $\mathbf{Z}_{kk},\mathbf{Z}_{bb}$ are the impedance matrices of $S_k$ and $S_b$ respectively, while $\mathbf{Z}_{kb},\mathbf{Z}_{bk}$ are their interaction matrices. Although convenient, this method requires discretization of the background structure.}. $\mathbf{\tilde{R}}$ is the real part of $\mathbf{\tilde{Z}}$, $\mathbf{J}_{k,n}$ is the expansion vector for the $n$-th characteristic current under Rao-Wilton-Glisson (RWG) basis functions \cite{ref_RWG}, and $\lambda_n$ is the corresponding characteristic number.

\begin{figure}[!t]
  \centering
  \includegraphics[]{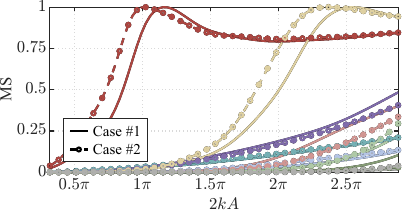}
  \caption{MS curves of SCM for the two models in Fig. \ref{fSheet_sphA} and \ref{fSheet_sphB}.}
  \label{fMS_sep}
\end{figure}

\begin{figure}[!t]
  \centering
  \includegraphics[]{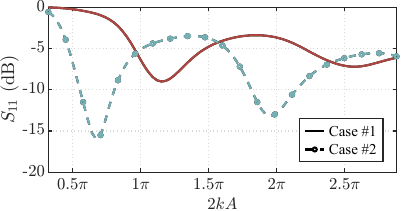}
  \caption{$S_{11}$ parameter for the two models in Fig. \ref{fSheet_sphA} and \ref{fSheet_sphB}, obtained using full-wave simulation software FEKO. The feeding is set on PEC sheet only.}
  \label{fS11}
  \vspace{-0.15in}
\end{figure}

Figure \ref{fMS_sep} illustrates the modal significance ($\mathrm{MS}=|\frac{1}{1+j\lambda_n}|$) of each SCM. Since $\mathrm{MS} = 1$ physically represents resonance, the correctness of these curves can also be corroborated by the S-parameter, as shown in Fig. \ref{fS11}. In the given frequency range, Fig. \ref{fMS_sep} and \ref{fS11} both indicate two resonance points at $2kA = 1.15 \pi$ and $2kA = 2.56 \pi$, corresponding to the main and the first higher-order mode, respectively. These results demonstrate the precise of SCMT in analyzing this type of structure.

Panning the suspended PEC sheet to just touch the background sphere yields case \#2, as illustrated in Fig. \ref{fSheet_sphB}. Repeating \eqref{eq1} leads to new MS curves represented by the dashed lines in Fig. \ref{fMS_sep}. These newly obtained MS curves are similar to those in case \#1, but with slightly shifted resonant frequencies. However, when compared with the S-parameter demonstrated by the dashed line in Fig. \ref{fS11}, significant differences are revealed.

\begin{figure}[!t]
  \centering
  \subfloat[]{\includegraphics[]{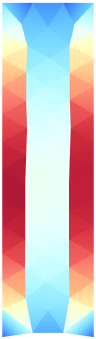}%
  \label{fcurr_sep4}}
  \hfil
  \subfloat[]{\includegraphics[]{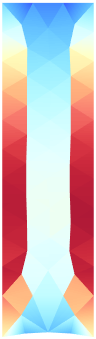}%
  \label{fcurr_sep0}}
  \hfil
  \subfloat[]{\includegraphics[]{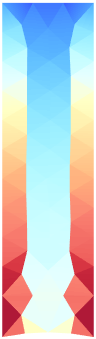}%
  \label{fcurr_fed}}
  \hfil
  \subfloat[]{\includegraphics[]{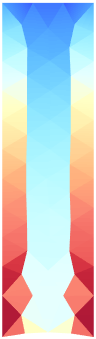}%
  \label{fcurr_connect}}
  \caption{Current distribution of: (a) main-mode in Fig. \ref{fMS_sep}, case \#1; (b) main-mode in Fig. \ref{fMS_sep}, case \#2; (c) full-wave simulation of case \#2; (d) main-mode in Fig. \ref{fMS_connect}, RCM.}
  \label{curr_distribute}
\end{figure}

By examining the current distribution on the PEC sheet, we found that the main-mode current in case \#2 (Fig. \ref{fcurr_sep0}) disappears at the boundary in contact with the background. However, the practical current should extend its path and flow to the background across the boundary due to the electrical connection. Therefore, the peak of the current should shift closer to the contact boundary like Fig. \ref{fcurr_fed}. Further comparison of the currents in case \#1 and \# 2 (Fig. \ref{fcurr_sep4} and \ref{fcurr_sep0}) indicates that the inaccuracy in case \#2 results from cutting off the current path from the PEC sheet to the background sphere (this conclusion is also supported by \cite{ref_Parhami_Technique}). This phenomenon occurs because discretizing a surface using RWG basis functions maintains current continuity within the surface but accumulates artificial line charges at its boundary. This accumulation causes a discontinuity in the current at the contact boundary and alters the current distribution on the PEC sheet.

\subsection{Eliminating Current Discontinuity at Contact Boundary}
\label{SecIIB}

\begin{figure}[!t]
  \centering
  \includegraphics[]{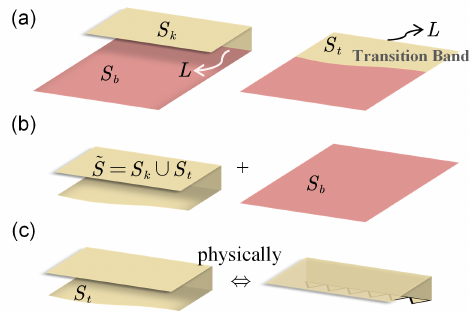}
  \caption{The concept of transition band. (a) original configuration and the defination of transition band $S_t$. Its EM solution is equivalent to (b), treating $S_k$ and $S_b$ as an entirety radiating against $S_b$. The transition band in (b) can also be reduced to a layer of serrate construction in (c).}
  \label{ftransband}
  \vspace{-0.15in}
\end{figure}

To eliminate the artificial current discontinuities caused by discretization, we introduce the concept of a transition band ($S_t$), illustrated in Figure \ref{ftransband}a. This transition band is a sub-surface of the background ($S_b$) that contains the contact boundary $L$ between the key structure ($S_k$) and the background. By incorporating this transition band, we redefine the structure as $\tilde{S}$, which consists of both $S_k$ and $S_t$.

Using $S_b$ as the radiation background for $\tilde{S}$, as shown in Fig. \ref{ftransband}b, we can solve for the characteristic current $\mathbf{\tilde{J}}_n$ by following the same procedure in \eqref{eq1}. Because $L$ is now an internal edge of $\tilde{S}$, the RWG basis functions defined on $\tilde{S}$ maintain current continuity across it, thus eliminating the artificial discontinuities.

Futher bifurcating $\mathbf{\tilde{J}}_n$ into subsets as
\begin{equation}
  \label{eq2}
  \tilde{\mathbf{J}}_n=\begin{bmatrix}
    \tilde{\mathbf{J}}_{k,n}& \tilde{\mathbf{J}}_{t,n}
  \end{bmatrix}^t
\end{equation}
where $\tilde{\mathbf{J}}_{k,n}$ and $\tilde{\mathbf{J}}_{t,n}$ represents the surface current on $S_k$ and $S_t$, respectively. It is important to note that $\tilde{\mathbf{J}}_{t,n}$ does not include any current component on $S_b$, despite $S_t\subset S_b$. This can be understood as a current near and tangent to $S_b$. Leveraging the reciprocity theorem \cite{ref_Harrington_EM_theory} and considering that $S_b$ is PEC, $\tilde{\mathbf{J}}_{t,n}$ generates no field elsewhere. Therefore, only $\tilde{\mathbf{J}}_{k,n}$ in \eqref{eq2} determines the EM properties (resonance and radiation) of $\tilde{S}$. 

Realizing this, we understand that $\tilde{\mathbf{J}}_{t,n}$ is introduced solely to ensure current continuity across the contact boundary $L$ and can thus be discarded. This is equivalent to removing $S_t$ from $\tilde{S}$. Therefore, $\tilde{\mathbf{J}}_{k,n}$ in deed represents the EM characteristics of $S_k$ and can be interpreted as the characteristic mode (i.e., RCM) of $S_k$, which will be further supported by numerical results in Sec. \ref{SecIIIA}.

The size of the transition band $S_t$ is theoretically arbitrary and primarily affects the scale of the matrices in \eqref{eq1} due to the inclusion of additional RWG basis functions. However, because the RWG basis functions involving only triangle pairs within $S_t$ generate zero fields, they introduce entire rows and columns of zero elements into the impedance matrix $\mathbf{\tilde{Z}}$, complicating the solution of \eqref{eq1}. These zero elements do not fundamentally impact the results of \eqref{eq1} and can be eliminated by removing these triangle pairs. Consequently, only a layer of serrated construction remains, representing RWG basis functions across $L$, with one triangle in $S_b$ (generates no fields) and the other in $S_k$ (generates fields), as illustrated in Fig. \ref{ftransband}c. This adjustment can be performed during the preprocessing stage of the model.

\section{Numerical Results}
\label{SecIII}
 
\subsection{Verification}
\label{SecIIIA}

\begin{figure}[!t]
  \centering
  \includegraphics[]{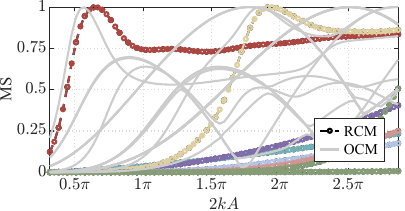}
  \caption{MS traces of RCM and OCM for the case \#2.}
  \label{fMS_connect}
\end{figure}

This section presents numerical results of case \#2 using the procedure in Sec. \ref{SecIIB}. The solid line in Fig. \ref{fMS_connect} demonstrates the MS curves of RCMs, where the two identified resonances occur at $2kA = 0.67\pi$ and $2kA = 1.95\pi$, aligning with the full-wave simulation results in Fig. \ref{fS11}. Unlike Fig. \ref{fMS_sep}, our method accurately reveals the resonant properties without inaccuracies. 

\begin{figure}[!t]
  \centering
  \includegraphics[]{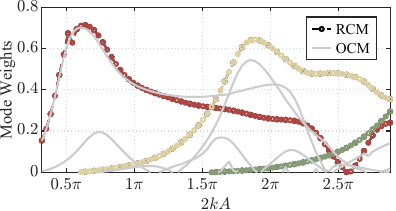}
  \caption{Mode weights of RCM and OCM for the case \#2.}
  \label{fMWC_connect}
  \vspace{-0.15in}
\end{figure}

\begin{figure}[!t]
  \centering
  \subfloat[]{\includegraphics[]{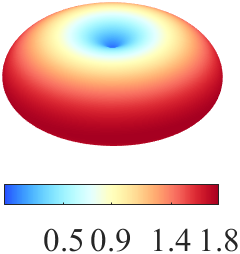}%
  \label{f_1Rad_RCM1}}
  \hfil
  \subfloat[]{\includegraphics[]{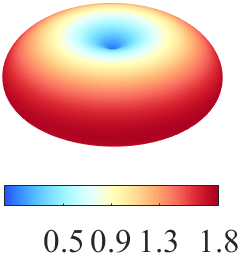}%
  \label{f_1Rad_fed1}}
  \hfil
  \subfloat[]{\includegraphics[]{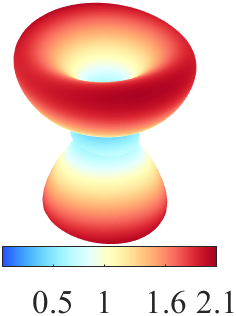}\
  \label{f_1Rad_RCM3}}
  \hfil
  \subfloat[]{\includegraphics[]{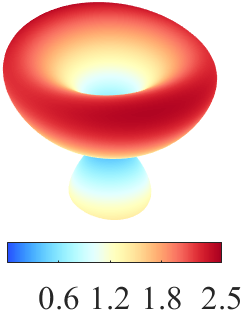}%
  \label{f_1Rad_fed3}}
  \caption{Radiation patterns. (a) and (c) are the results of the first RCM @ $2kL=0.67\pi$ and the second @ $2kL=1.95\pi$, respectively. (b) and (d) are full-wave simulation results at the corresponding frequencies.}
  \label{f_1Rad}
  \vspace{-0.15in}
\end{figure}

Further validation is observed by comparing the RCM currents and radiation patterns with full-wave results. Since the main-mode component dominants completely at $2kA = 0.67\pi$, as illustrated in Fig. \ref{fMWC_connect}, figures \ref{fcurr_fed} and \ref{fcurr_connect}, \ref{f_1Rad_RCM1} and \ref{f_1Rad_fed1} are almost identical. Although Fig. \ref{f_1Rad_RCM3} and \ref{f_1Rad_fed3} show slight differences, this is due to the coexistence of the first two RCMs at $2kA = 1.95\pi$. 

\begin{figure}[!t]
  \subfloat[]{
    \begin{minipage}{\linewidth}
    \centering
    \includegraphics[height=1.05in]{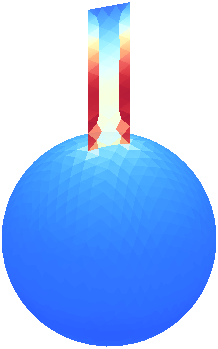}
    \hfil
    \includegraphics[height=1.05in]{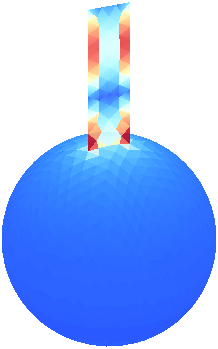}
    \hfil
    \includegraphics[height=1.05in]{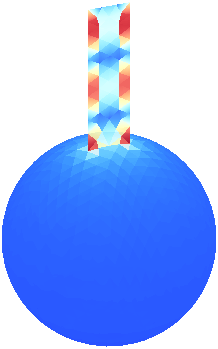}
  \end{minipage}
  }
  \vfil
  \vspace{-0.15in}
  \subfloat[]{
    \begin{minipage}[]{\linewidth}
    \centering
    \includegraphics[height=1.05in]{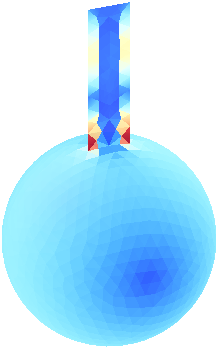}
    \hfil
    \includegraphics[height=1.05in]{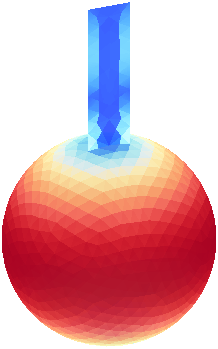}
    \hfil
    \includegraphics[height=1.05in]{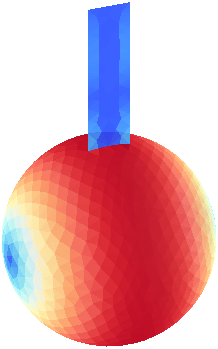}
  \end{minipage}
  }
  \vfil
  \vspace{-0.15in}
  \subfloat[]{
    \begin{minipage}{\linewidth} 
    \centering
    \includegraphics[]{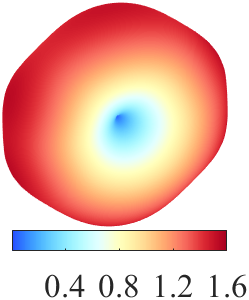}
    \hfil
    \includegraphics[]{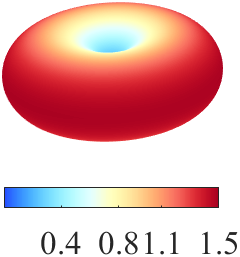}
    \hfil
    \includegraphics[]{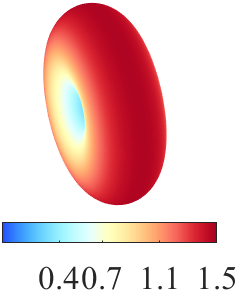}
  \end{minipage}
  }
  \caption{Current distribution for the first three (a) RCMs and (b) OCMs at $2kA=1.95\pi$. (c) is the radiation patterns corresponding to (b).}
   \label{fRCMvsOCM}
\end{figure}

The RCM also exhibits advantages over the original CM (OCM). As shown in Fig. \ref{fMS_connect}, the MS traces of OCMs are cluttered due to the inclusion of many direct radiation components from the sphere background. Most OCM currents concentrate on the sphere rather than the PEC sheet, and their corresponding radiation patterns always exhibit upper-and-lower symmetry, as demonstrated in Fig. \ref{fRCMvsOCM}. Consequently, none of the OCMs can independently support the design of the PEC sheet near a metal sphere background, even at the main-mode frequency, as shown in Fig. \ref{fMWC_connect}.

In fact, the additional engaged OCMs can be viewed as synthesizing the RCM, as their background direct radiation effects counteract each other. However, RCM analysis inherently pre-removes the background's direct radiation effect, thereby facilitating the characterization of the structure's radiation using fewer modes, or even just a single mode. This makes the RCM a more efficient and effective tool for analyzing and designing antenna structures in complex environments.

\subsection{Application}
\label{SecIIIB}

\begin{figure}[!t]
  \centering
  \begin{tikzpicture}
    \node (img1) {\includegraphics{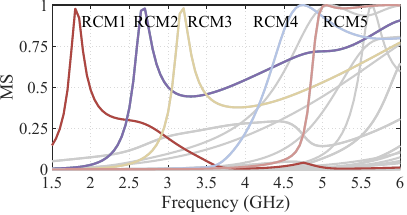}};
    \node at (img1.center)[shift={(3.4cm,0.1cm)}]{\includegraphics{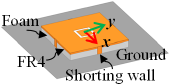}};
  \end{tikzpicture}
  \caption{MS traces of the MPA configuration reported in \cite{ref_MPA}.}
  \label{fMPA_TAP}
\end{figure}

Utilizing our validated RCM method, we set out to analyze and improve an existing dual-band MPA antenna. The prototype of the MPA antenna is shown in Fig. \ref{fMPA_TAP}, with its dimensions detailed in \cite{ref_MPA}. Our analysis diverges from \cite{ref_MPA} only in the removal of the antenna's loss. Note that this structure features a shorting wall connecting the ground to the top patch.

Focusing on the top two metal patches of the MPA as the key substructure, we conducted RCM analysis, results of which are depicted in Fig. \ref{fMPA_TAP}. In the frequency range of 1.5 GHz to 6 GHz, we identified a significantly higher number of modes than the two (resonating at 2 GHz and 4.9 GHz) originally reported in \cite{ref_MPA}.

\begin{figure}[!t]
  \centering
  \subfloat[]{\includegraphics[]{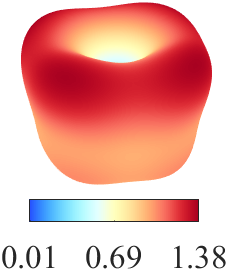}%
  \label{fpat_MPAAP_CM1}}
  \hfil
  \subfloat[]{\includegraphics[]{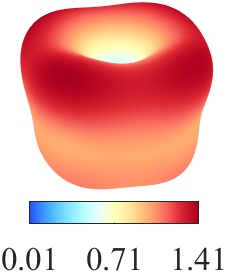}%
  \label{fpat_MPAAP_HFSS1}}
  \hfil
  \subfloat[]{\includegraphics[]{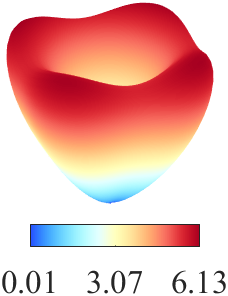}%
  \label{fpat_MPAAP_CM2}}
  \hfil
  \subfloat[]{\includegraphics[]{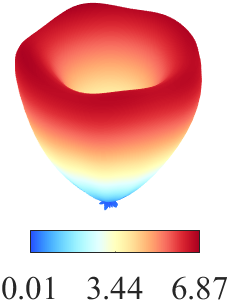}%
  \label{fpat_MPAAP_HFSS2}}
  \caption{Radiation patterns of (a) RCM1, (b) mode 1 in \cite{ref_MPA}, (c) RCM4, and (d) mode 2 in \cite{ref_MPA}. The results of \cite{ref_MPA} were re-simulated using HFSS. Due to the existence of the main-mode in (d), it differs from (c) moderately.}
  \label{fpat_MPAAP}
  \vspace{-0.15in}
\end{figure}

Further analysis of the RCMs indicates that the two modes previously identified through the S-parameter in \cite{ref_MPA} precisely correspond to RCM1 and RCM4, respectively resonating at ~{1.9 GHz} and 4.8 GHz. This finding is corroborated by the radiation patterns of these modes as illustrated in Fig. \ref{fpat_MPAAP}, which align with the results presented in \cite{ref_MPA}.

\begin{figure}[!t]
  \centering
  \subfloat[]{\includegraphics[height=0.65in]{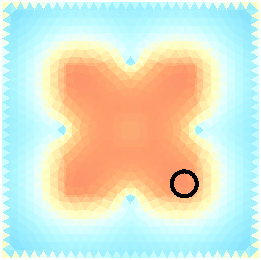}%
  \label{RCM1}}
  \hfil
  \subfloat[]{\includegraphics[height=0.65in]{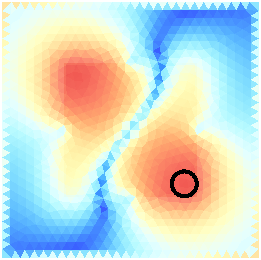}%
  \label{RCM2}}
  \hfil
  \subfloat[]{\includegraphics[height=0.65in]{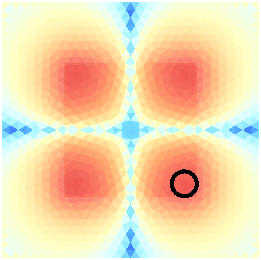}%
  \label{RCM3}}
  \hfil
  \subfloat[]{\includegraphics[height=0.65in]{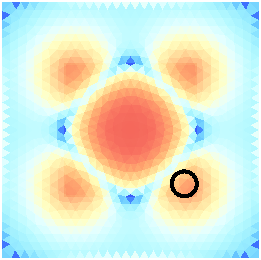}%
  \label{RCM4}}
  \hfil
  \subfloat[]{\includegraphics[height=0.65in]{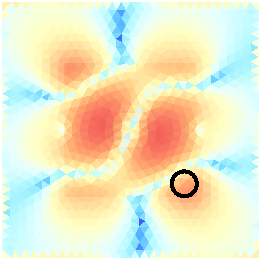}%
  \label{RCM5}}
  \caption{Electric field distributions for: (a) RCM1, (b) RCM2, (c) RCM3, (d) RCM4, and (e) RCM5. The location of circle mark is $x=y=21$mm.}
  \label{fE_fdisb}
\end{figure}

A detailed examination of the RCMs' electric field (E-field) distributions sheds light on why the MPA is operational only at these two frequency bands. The excitation probe in the MPA is located at the center, where most RCMs show minimal E-field intensities, with the notable exceptions of RCM1 and RCM4, as depicted in Fig. \ref{fE_fdisb}. Therefore, only RCM1 and RCM4 are effectively excited and observable in \cite{ref_MPA}.

\begin{figure}[!t]
  \centering
  \includegraphics[]{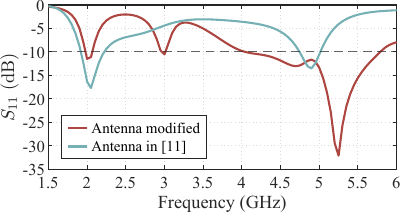}
  \caption{$S_{11}$ parameter of the MPA.}
  \label{fS11_MPA_TAP}
\end{figure}

Following our analysis, we propose enhancing the MPA's operational frequency bands and bandwidth by modifying its feed position to engage more modes. As indicated in Fig. \ref{fE_fdisb}, we repositioned the excitation probe to the corner ($x=y=21$mm), where E-field intensities are substantial across modes RCM1 to RCM5. This modification, coupled with adjusting the characteristic impedance of the excitation port to 85 ohms, resulted in three operational bands, marked by the $S_{11}$ curve (red line) in Fig. \ref{fS11_MPA_TAP}. The three bands achieved -10dB relative bandwidths of 4.5\% at 2 GHz, 1.4\% at 3 GHz, and 35.4\% at 5.25 GHz, respectively. By examining the input impedance, we found that the first band originates from RCM1, the second from a combination of RCM2 and RCM3, and the third from a blend of RCM4 and RCM5. Compared to the design in \cite{ref_MPA}, we have increased the high-frequency bandwidth by 30.1\%. Note that the reduction to the low-frequency bandwidth is due to the anti-resonant properties of RCM2, which leads to steep changes in the low-frequency impedance.

\section{Conclusion}

This article introduces a transition band method to eliminate current discontinuity in computing the SCMs of interconnected metallic structures. The accuracy of this method has been validated through comparisons with full-wave simulation features. Furthermore, the practicality of our method is demonstrated through a case study, where employing RCM analysis substantially improves the performance of an existing MPA. This innovative method extends the application range of SCMT and offers a valuable tool for the precise analysis of antennas that include interconnected multi-metal structures.

\vfill

\end{document}